%% file: main.tex
\documentclass[12pt]{article}
\input{preamble}

\graphicspath{{./figures/}}%

\begin{document}

\title{\vspace{-1em}\Large\bf A Response to Recent Critiques of  Hainmueller, Mummolo and Xu (2019) on Estimating Conditional Relationships%
\thanks{Jens Hainmueller, Professor of Political Science, Stanford University. Email: \url{jhain@stanford.edu}. Jiehan Liu, PhD student, Department of Political Science, Stanford University. Email: \url{jiehanl@stanford.edu}. Ziyi Liu, PhD student, Haas School of Business, University of California, Berkeley. Email: \url{zyliu2023@berkeley.edu}. Jonathan Mummolo, Associate Professor of Politics and Public Affairs, Princeton University. Email: \url{jmummolo@princeton.edu}. Yiqing Xu, Associate Professor of Political Science, Stanford University. Email: \url{yiqingxu@stanford.edu}. This response draws heavily on \citet{LLX}. We thank Chad Hazlett and Dean Knox for helpful comments and suggestions. This version corrects an error in the February 2025 version: for a continuous treatment and no additional covariates, the kernel estimator recovers the CME if the outcome is linear in treatment at each moderator value, and for every outcome function only if treatment is normally distributed conditional on the moderator. We thank Josep Serrano-Serrat for pointing it out \citep{serranoserrat2026kernel}. The authors used Claude Code (Anthropic) for independent checks of the mathematics and for proofreading. All errors are our own.}
}

\author{\normalsize Jens Hainmueller\\ \normalsize (Stanford)\and \normalsize Jiehan Liu\\ \normalsize (Stanford)
\and \normalsize Ziyi Liu\\ \normalsize (Berkeley) \and \normalsize Jonathan Mummolo\\ \normalsize (Princeton)\and \normalsize Yiqing Xu\\ \normalsize (Stanford)}

\date{}

{\setlength{\tabcolsep}{2pt}\maketitle} %

\vspace{-1.5em}
\begin{abstract}
\noindent \cite{simonsohn2024blog} and \cite{simonsohn2024interacting} critique \citet[][HMX]{HMX2019}, arguing that failing to model nonlinear relationships among the treatment, the moderator, and the outcome leads to biased marginal effect estimates and uncontrolled Type-I error rates. While these critiques highlight the issue of under-modeling nonlinearity in applied research, they are flawed in several key ways. First, the causal estimand for interaction effects and the necessary identifying assumptions are not clearly defined in these critiques. Once properly stated, the critique of the key example no longer holds. Second, the kernel estimator of HMX recovers the true causal effects in the key simulated example of the recent critiques, which benchmarked the estimators against the slope at a fixed treatment level instead of the conditional marginal effect (CME), and thus arrived at misleading conclusions. Third, while HMX acknowledge that Generalized Additive Models (GAM) can be a useful exploratory tool, the GAMs used in these critiques estimate the effect at a chosen treatment level rather than the CME, and better alternatives exist, particularly in the presence of additional covariates. Our response aims to clarify these misconceptions and provide updated recommendations for researchers studying interaction effects through the estimation of the CME.


\end{abstract}

\newpage
\doublespace

\setcounter{page}{1}
\abovedisplayskip=5pt
\belowdisplayskip=5pt

\noindent A recent blog post \citep{simonsohn2024blog} and accompanying paper \citep{simonsohn2024interacting} critique a paper published in \textit{Political Analysis} titled ``How Much Should We Trust Estimates from Multiplicative Interaction Models? Simple Tools to Improve Empirical Practice'' \citep[][hereafter HMX 2019]{HMX2019}. These critiques argue that the binning and kernel estimators proposed in HMX (2019) are inappropriate for estimating and testing conditional relationships using observational data. 

We, the authors of HMX (2019), have appreciated these critiques and shared our initial response in private. However, since some important substantive points were not reflected in the blog post, we (HMX and two additional contributors) have decided to provide the following public response. We believe this will help clarify the issues raised and contribute to improving research practices.

\section*{Critiques of HMX (2019)}

Let $Y$, $D$, $X$, and $Z$ denote the outcome, treatment, moderator, and a set of additional covariates, respectively. Empirical researchers are often interested in how the effect of treatment $D$ on outcome $Y$ varies with a covariate $X$, which we call the moderator. There may exist other covariates $Z$, and we denote $V = (X, Z)$. While HMX (2019) also considers panel settings, for simplicity, we focus here on a cross-sectional setting where $(Y_{i}, D_{i}, V_{i})$ are i.i.d. from a super-population. Traditionally, researchers often estimate the following linear interaction model:
\begin{equation}\label{mod1}
Y = \beta_{0} + \beta_{1} D + \beta_{2} X + \beta_{3} DX + \gamma Z + \epsilon,
\end{equation}
where $\epsilon$ represents an idiosyncratic error, and interpret $\widehat{ME}(x) = \hat\beta_{1} + \hat\beta_{3}x$ as the conditional marginal effect of $D$ on $Y$ with respect to $X$ \citep{BCG2006}. 

HMX (2019) highlights that Model~(\ref{mod1}) is often unrealistic because this functional form implies that the effect of $D$ on $Y$ changes linearly with $X$, and the overlap assumption is frequently violated. To address these challenges, HMX (2019) recommends (i) visual diagnostics such as subgroup scatterplots, (ii) the \textbf{Generalized Additive Models (GAM) plot}, and (iii) the binning estimator, as diagnostic tools. Additionally, HMX (2019) proposes a semiparametric kernel estimator, implemented via local linear regressions, to relax the functional form assumption:
\begin{equation}\label{mod2}
Y = f(X) + g(X) D + \gamma(X) Z + \epsilon.
\end{equation}
Hence, the conditional marginal effects are estimated with a more flexible functional form which can vary freely across levels of $X$: $\widehat{ME}(x) =  \hat{g}(x)$.

The primary concern in \cite{simonsohn2024blog} and \cite{simonsohn2024interacting} is the misspecification bias that arises from nonlinear relationships between $D$, $X$, and $Y$, particularly when $D$ and $X$ are correlated. In these analyses---both in the paper and the blog---no additional covariates $Z$ are included. As noted in  \cite{simonsohn2024interacting}, this bias increases the likelihood of detecting spurious interactions, leading to an elevated false positive rate. To account for nonlinearity, GAMs of the following form are recommended:%
\[
Y = f(D, X) + \epsilon.%
\footnote{Specifically, \citet{simonsohn2024interacting} advocates for a model incorporating three smooth functions: $$Y = f_{1}(D) + f_{2}(X) + f_{3}(D, X) + \epsilon.$$
The smooths are penalized, and the tensor-product term can have separate penalties for its two directions. The term $f_{3}(D, X)$ is intended to capture the interaction effect. It is implemented using  \texttt{gam(y $\sim$ s(x1) + s(x2) + ti(x1, x2), data = foo)} with the \texttt{mgcv} package in \texttt{R}.}
\]

\begin{figure}[!ht]
    \caption{Figure adapted from \citet{simonsohn2024blog}\\
    (with Modified Labels)}
    \begin{center}
    \includegraphics[width=0.7\linewidth]{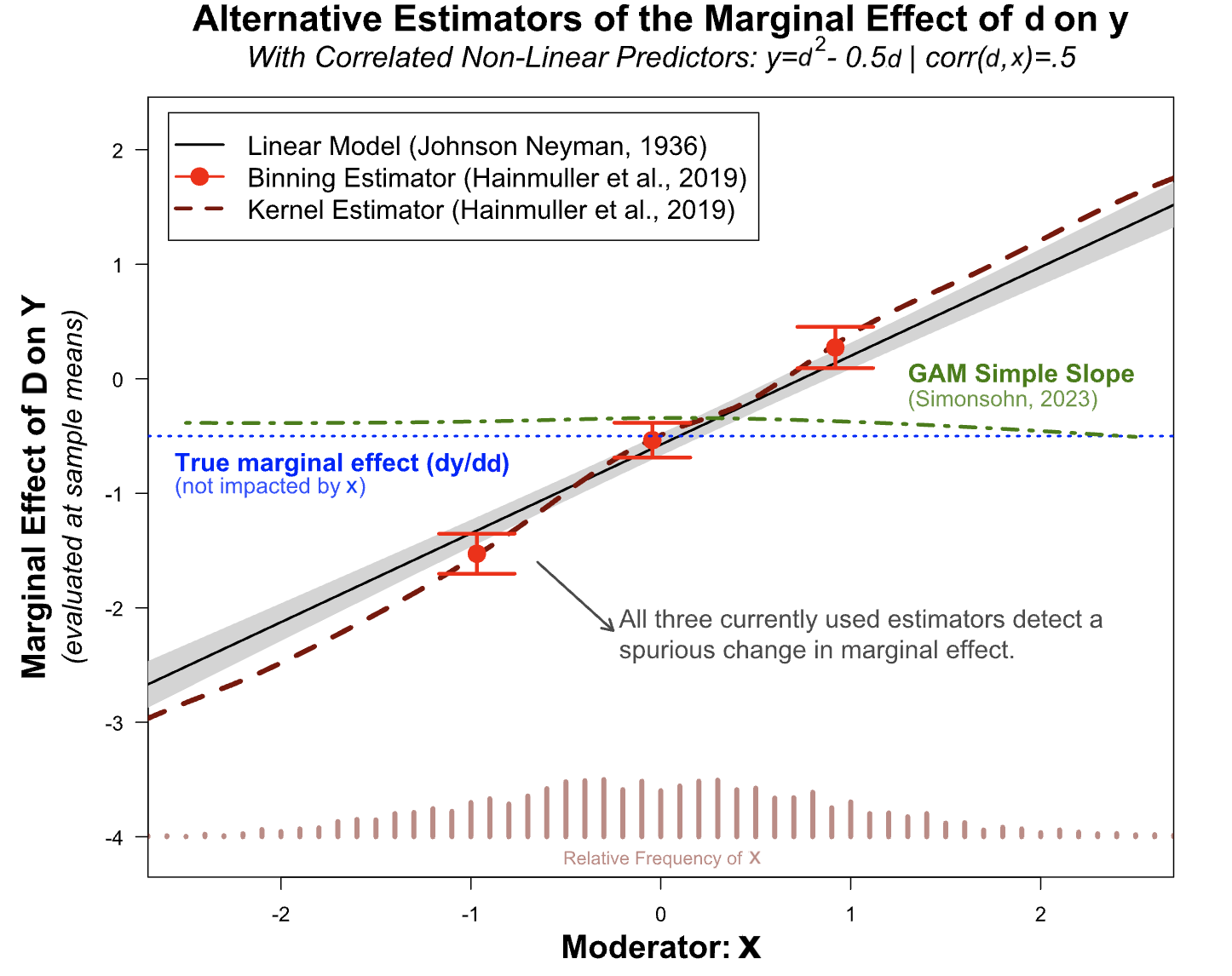}
    \label{fig:sim}\vspace{-1em}
    \end{center}
    \textbf{Note:} In the original plot, the treatment and moderator were labeled as $X$ and $Z$, respectively; we have modified them to $D$ and $X$ to maintain consistency with the notation used in this response.
\end{figure}

\citet{simonsohn2024blog} provides an intriguing simulated sample with the following data-generating process (DGP): $Y = D^2 - 0.5D + \epsilon$, where $X$ and $D$ are correlated: $Corr(X, D) = 0.5$. The author claims that the marginal effect of $D$ on $Y$ conditional on $X$ does not change with $X$, i.e., that $\partial ME(x)/\partial x = 0$, while the linear estimator, the binning estimator, and the kernel estimator all produce marginal effect estimators that are linearly and positively correlated with $x$ (Figure~\ref{fig:sim}). Based on the shared \texttt{R} code, we believe that the estimation actually targets a finite-difference approximation to $\mathrm{E}\left[\left.\frac{\partial Y}{\partial D} \right\rvert\, D = 0, X\right]$.%
\footnote{The shared code is provided in the footnote of Figure 3 in \citet{simonsohn2024blog}. The procedure begins by fitting a GAM: \texttt{gam(y $\sim$ s(D) + s(X) + ti(D, X), data = foo)}. Then, the predicted values of $Y$ are computed over a range of $X$ values while holding $D$ fixed at 0 and 0.1. The ``marginal effects" $\partial Y /\partial D$ are then estimated using a numerical approximation-specifically, by taking the difference between the predicted $Y$ values at $D=0.1$ and $D=0$, and dividing by 0.1. A similar procedure for GAM simple slopes is advocated in \citet{simonsohn2024interacting}.} However, this approach does \emph{not} align with the statistical definition of marginal effects (see further discussion below). 

The disagreement may also stem from a difference in understanding of the word ``marginal.'' In \citet{simonsohn2024blog}, a ``marginal effect'' is a partial derivative with respect to the treatment: the slope of the outcome in $D$ at a given treatment level $d$ and moderator value $x$, $\partial \E[Y \mid D = d, X = x]/\partial d$. In this response and in \citet{LLX}, following the statistical tradition, ``marginal'' means that the expectation is taken over the distribution of the treatment and the other covariates at each value of $X$. The two coincide in the linear interaction model but differ when the outcome is nonlinear in $D$, as in this example \citep[Section 1.3]{LLX}. We define both quantities formally in the next section.

\paragraph*{Merits of the Critique.} There is value in these critiques. First, they bring renewed attention to the issue of potential nonlinearity in estimating and testing conditional relationships. Although the kernel estimator can recover the intended estimand in the key example (as we will show below), it requires a large amount of data when nonlinear terms of $D$ are present, and with a continuous treatment it does so only under additional conditions. Second, these critiques present thought-provoking examples that challenge existing methods. Although they are flawed in several key ways, they demonstrate the importance of clearly defining estimands and often-unstated identifying assumptions.

\paragraph*{Summary of our response.} We address these critiques with three main points:
\begin{enumerate}\itemsep0em
    \item The most critical flaw in the critique is the lack of a clear definition of the estimand (and identifying assumptions). The critique relies on GAM prediction curves while simultaneously using strong causal language, such as ``effect'' or ``marginal effect.'' We will show that once these elements are clearly defined, the critique of the key example no longer holds---in fact, the GAM output, evaluated at a chosen level of the treatment, approximates the conditional average partial effect (CAPE) at that level rather than the CME.
    \item Once the estimand, the conditional marginal effect (CME), and key assumptions are clearly stated, existing methods can estimate it without assuming that the effect is linear in $X$. These methods include the kernel estimator (with sufficient data, an appropriately small bandwidth, and, for a continuous treatment, an outcome that is linear in $D$ at each value of $X$ or a treatment that is normal given $X$), augmented inverse probability weighting (AIPW), partialing-out Lasso (PO-Lasso), and double/debiased machine learning (DML) estimators. AIPW and DML with a binary treatment are doubly robust, and all three are Neyman orthogonal, making them more resilient to model misspecification in applied settings. With a continuous treatment, the PO-Lasso and DML estimators in \citet{LLX} likewise assume an outcome linear in $D$ with a slope $\theta(X)$ common across $Z$; otherwise they recover a variance-weighted linear projection that need not equal the CME.        
    \item The method advocated in \citet{simonsohn2024blog} and \citet{simonsohn2024interacting} is GAM. As used there, it approximates the CAPE at a chosen treatment level, not the CME. The CME can be obtained from a GAM fit by averaging each unit's fitted slope at its own treatment level \citep{serranoserrat2026kernel}, but it relies heavily on researchers' prior knowledge of where nonlinearity may occur---a requirement that is often unrealistic in applied settings, especially when dealing with high-dimensional covariates \(Z\). This limitation makes GAM susceptible to both regularization and misspecification biases. GAM is a useful exploratory tool and, used this way, a reasonable estimator of the CME when there are few covariates.
\end{enumerate}
\bigskip

\section*{Properly Defining the Estimand Is Critical}

First, we highlight the critical importance of clearly defining the causal estimand, which is often neglected in social science research \citep{lundberg2021your}. HMX (2019), targeting an audience familiar with the regression framework, does not formally adopt modern causal inference notation; we address this limitation in this response. We show that once the estimand, the CME, is properly specified, the conclusion \citet{simonsohn2024blog} draws from its key example no longer holds. To illustrate this, we revisit the key simulated example from \citet{simonsohn2024blog}, demonstrating that its misleading conclusion stems from a misunderstanding of the estimand. 

We begin by defining the relevant estimands and outlining the key identifying assumptions within the Neyman-Rubin potential outcomes framework \citep{neyman1923application, rubin1974estimating}.
  
\begin{assumption} (SUTVA).\label{assm:sutva}
Potential outcome of unit $i$ can be written as:
\begin{center}
$Y_i(d_{1}, d_{2}, \cdots, d_{n}) = Y_i(d_{i})$
\end{center}
in which $d_{i}$ denote the potential treatment value for unit $i$. Therefore, $Y_{i} = Y_i(D_{i})$, which connects the observed outcome with potential outcomes. 
\end{assumption}
Given SUTVA, we define the main estimand, the conditional marginal effect (CME): 
\begin{equation*}
\label{eq:CME}
\theta(x) = \E\left[ \left.\frac{\partial Y_i(d)}{\partial d}\right|_{d = D_i} \Bigg| X_i = x \right].
\end{equation*}
It is \emph{causal} because it represents an aggregation of comparisons between potential outcomes for the same unit. When $D_{i}$ is binary, it is defined as
\begin{equation*}
\theta(x) = \E[Y_i(1) - Y_i(0) \mid X_i = x].
\end{equation*}
which is essentially conditional average treatment effect (CATE) at covariate value $v$, 
$\tau(v) = \E[Y_i(1) - Y_i(0) \mid V_i = v]$, \emph{marginalized} over the distribution of additional covariates $Z_{i}$. 

It is important to note that the CME does \emph{not} imply that the CATE changes with $X$ in a causal manner. The concept of \emph{causal moderation} (also known as causal interaction), by contrast, refers to the causal effect of $X$ on the treatment effect of $D$ on $Y$ \citep{tyler, bansak2020estimating}. To define causal moderation, we need to extend the potential outcome framework to include $x$ explicitly and define causal moderation at $x$ as:
\[
\delta(x) = \E\left[ \left.\frac{\partial^2 Y_i(d, x)}{\partial d \partial x}\right|_{d = D_i} \Bigg| X_i = x \right],
\]
where the expectation is taken over $D$ and $Z$. Identifying $\delta(x)$ requires much stronger assumptions, including the quasi-randomization of $X$, which are far more stringent than those needed for identifying the CATE or CME.

Another estimand, often conflated with the CME, is the \emph{conditional average partial effect} (CAPE):  
\[
\rho(d, x, z) =  \E\left[ \frac{\partial Y_i(d)}{\partial d} \Bigg| D_i = d, X_i = x, Z_i = z \right],
\]
which represents the average partial effect of $D$ on $Y$ given specific values of $D$, $X$, and $Z$. When $D$ is binary, it simplifies to $\E[Y_i(1) - Y_i(0) \mid D_i = d, V_i = v], d = \{0, 1\}$, which under unconfoundedness equals $\tau(v)$ for either treatment group. This estimand differs from the conventional interpretation of marginal effects in that (i) the influence of $Z$ is not marginalized; (ii) it is conditional on a specific $d$. Therefore, we rarely see this estimand used in empirical work. However, as we will see below, this may be the implicit estimand in \cite{simonsohn2024blog}.

\begin{table}[!ht]
\centering
\caption{Summary of Estimands}
\label{table:estimands}
\begin{tabular}{@{} >{\raggedright\arraybackslash}p{0.25\textwidth} 
                 >{\raggedright\arraybackslash}p{0.35\textwidth} 
                 >{\raggedright\arraybackslash}p{0.35\textwidth} @{}}
\toprule
\textbf{Estimand} & \textbf{Definition} & \textbf{Interpretation} \\
\midrule
\small Conditional average treatment effect (CATE) & 
$\mathbb{E}\left[ \left.\frac{\partial Y_i(d)}{\partial d}\right|_{d = D_i} \big| X_i = x, Z_{i} = z \right]$ & 
\small The average effect of a treatment on a subgroup defined by $(X_i = x, Z_i = z$), marginalized over different values of $D$\\
\addlinespace
\small Conditional average partial effect (CAPE) & 
$\mathbb{E}\left[ \frac{\partial Y_i(d)}{\partial d} \big| D_i = d, X_i = x, Z_i = z \right]$ & 
\small CATE conditional on a specific value of $D$ (rarely used in continuous $D$ settings) \\
\addlinespace
\small Conditional marginal effect (CME) & 
$\mathbb{E}\left[ \left.\frac{\partial Y_i(d)}{\partial d}\right|_{d = D_i} \big| X_i = x \right]$ & 
\small The CATE averaged over $Z \mid X = x$; for continuous $D$, the CATE already averages slopes over $D \mid X, Z$.\\
\addlinespace
\small Causal moderation & 
$\mathbb{E}\left[ \left.\frac{\partial^2 Y_i(d, x)}{\partial d \partial x}\right|_{d = D_i} \big| X_i = x \right]$ & 
\small Also known as causal interaction effect; marginalized over $D$ and~$Z$. \\
\bottomrule\\
\end{tabular}
\end{table}

Table~\ref{table:estimands} summarizes these estimands, each distinct yet closely related. CATE measures the average effect of a change in treatment $D$ on the outcome $Y$, among units characterized by $(X_i = x, Z_i = z)$. This estimand is typically explored when researchers are interested in the heterogeneity of treatment effect. The CAPE additionally conditions on a specific treatment level $D_i = d$; it is rarely the target in continuous treatment settings. 
CME measures how the effect of treatment $D$ on outcome $Y$ varies across units with characteristic $X$, averaging over other characteristics $Z$ and the treatment level. In applied contexts, CME is frequently referred to as an interaction effect, highlighting how the treatment effect varies across levels of the moderator $X$, while disregarding other covariates. Lastly, causal moderation examines the second derivative with respect to both $D_i = d$ and $X_i =x$. Instead of asking how the effect of $D$ on $Y$ changes with the value of $X$ \emph{descriptively}, causal moderation examines how $X$ \emph{causally} impacts the effect of $D$ on $Y$; therefore, its identification typically requires (quasi-) randomization of $X$. 

Conventionally, researchers use linear interaction models to estimate the CME, which captures how the causal effect of $D$ on $Y$ varies with the value of $X$ (marginalizing over the distribution of $D$ and $Z$ at that value of $X$, hence the term ``marginal effects"). However, recent critiques of HMX (2019) lack clarity regarding the estimand, causing confusion. These critiques may potentially target either the CAPE (according to the shared \texttt{R} code) or causal moderation (according to the narrative, which asks whether the moderator changes the effect of the treatment). The former being an unconventional estimand and the latter requiring very strong identifying assumptions, such as (quasi-)randomization of both $D$ and $X$.

\paragraph*{The key simulated example.} Now, we examine the key simulated example through the lens of the CME. Consistent with the notation in HMX (2019), $D$ represents the treatment, and $X$ denotes the moderator. Note that \citet{simonsohn2024blog} uses $X$ and $Z$ to represent the treatment and moderator, respectively. From the simulation setup, we know the data-generating process (DGP) is as follows:  
\[
Y \;=\; D^2 \;-\; 0.5\,D  \;+\; \varepsilon,
\]
in which $\E[\varepsilon \mid D,X] = 0$ (with $Y_i(d) = d^2 - 0.5d + \varepsilon_i$ at every treatment level $d$, so $\partial Y_i(d)/\partial d = 2d - 0.5$) and 
\[
(D, X) \sim \mathcal{N}\left( \mathbf{0}, \begin{bmatrix} 1 & 0.5 \\ 0.5 & 1 \end{bmatrix} \right).
\]
The normality assumption gives us closed-form expressions for $\E[D \mid X=x]$ and the true CME. In Appendix Section A.1, we show that the true CME is
\[
\theta(x) = \E\!\Bigl[\frac{\partial Y}{\partial D} \Bigm| X=x\Bigr] =
x - 0.5.
\]
That is, if we hold $X=x$ fixed, the average slope of $Y$ with respect to $D$ is $(x - 0.5)$. Unlike what \cite{simonsohn2024blog} implies, the true CME is \emph{not} flat: it increases monotonically with $x$. The intuition is that as $X$ increases, the distribution of $D$ also shifts upward. Because the partial effect of $D$ on $Y$ is increasing in $D$, the CME also increases, even though $X$ does not directly enter the outcome equation.

In Appendix Section A.1, we prove that the linear estimator is asymptotically biased for the CME, whereas the kernel estimator proposed in HMX (2019) can accurately recover the CME in this simulated example. The reason is specific to this example. Near $X = x_0$, the kernel estimate converges to the slope of the best linear projection of $Y$ on $D$. This projection is the straight line that best approximates the outcome in $D$ among units with that value of $X$. The projection slope averages the slope $2D - 0.5$ with weights that favor values of $D$ near the center of its distribution \citep{yitzhaki1996using}. Because $D$ is normal given $X$, these weights equal the density of $D$ given $X$, so the projection slope equals the CME. The binning estimator will be biased but, in this example, has much less bias than the linear estimator at the low and high bin evaluation points, as it is a (rectangular) kernel estimator with a large bandwidth (bin size).

\begin{figure}[!h]
    \caption{Estimated CME on the Simulated Example} \label{fig:simu_toy_D}
    \centering
    \begin{minipage}{0.95\linewidth}
    \begin{center}
    \includegraphics[width=1\linewidth]{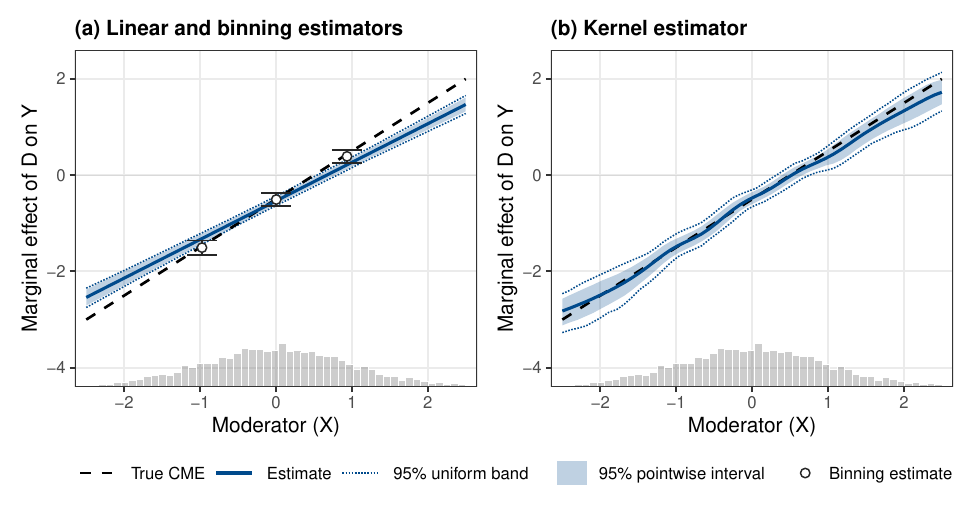}
    \end{center}\vspace{-1em}
    {\footnotesize \textbf{Note:} The dashed line is the true CME and the solid line is the estimate. Shaded areas are 95\% pointwise confidence intervals and dotted lines are 95\% uniform confidence bands. In panel (a), circles with bars are the binning estimates and their 95\% confidence intervals. The histogram at the bottom shows the distribution of $X$. The sample size is 5,000. The figure covers $X$ between $-2.5$ and $2.5$, as in Figure~\ref{fig:sim}.}
    \end{minipage}    
\end{figure}

Interestingly, these results are presented in \citet{simonsohn2024blog} (see Figure~\ref{fig:sim}); however, \citet{simonsohn2024blog} misinterprets them as evidence that the kernel (and binning) estimators are inherently biased. This misinterpretation arises because he conflates the CME, $\theta(x) = \mathbb{E}\left[ \left.\frac{\partial Y_i(d)}{\partial d}\right|_{d = D_i} \big| X_i = x \right]$, with the CAPE,  $\rho(d, x) =  \E\left[ \frac{\partial Y_i(d)}{\partial d} \mid D_i = d, X_i = x\right]$. We replicate the results in Figure~\ref{fig:simu_toy_D}, where the dashed line represents the true CME. As expected, the left panel shows that the linear estimator is biased, whereas the right panel demonstrates that, far from being ``false,'' the kernel estimator can recover the true CME fairly well in finite samples ($n = 5,000$), even when $Y$ depends on $D$ in a quadratic form and $X$ and $D$ are correlated. Specifically, in this draw the true CME lies within the 95\% uniform confidence band from the kernel estimator. 

As for the binning estimator, HMX (2019) proposes it as a diagnostic tool and develops a test based on it that compares the linear interaction model with a model allowing separate coefficients by bin. Of course, the binning estimator is a crude way to approximate the CME, as few researchers would actually assume that the CME is piecewise linear and only varies by terciles of $X$. Because the binning estimator uses data beyond a small neighborhood of the evaluation point $x_{0}$ when estimating the CME, it is likely biased if the relationship between $D$ and $Y$ is nonlinear at $x_{0}$. Nevertheless, as shown in Figure~\ref{fig:simu_toy_D}, the estimates produced by the binning estimator remain fairly close to the true CME.

The kernel estimator recovers the CME in the key example, but not in general. Consider the DGP used in \citet{simonsohn2024interacting}, $Y = f(D,X) + \epsilon$, where $f$ can be nonlinear in a continuous $D$. The kernel estimator fits a straight line in $D$ at each value of $X$. At $X = x$, this line converges to the best linear projection of $Y$ on $D$ among units with that value of $X$. Either of two conditions guarantees that the projection slope equals the CME. The first is that $f$ is linear in $D$ at that value of $X$, whatever the distribution of $D$. A binary treatment always meets this condition, since any function of a binary $D$ is linear in $D$. The second is that $D$ is normally distributed given $X$, as in the key example, whatever the shape of $f$. Normality is also necessary if the equality is to hold for every sufficiently regular $f$ \citep{serranoserrat2026kernel}. When neither condition holds, the projection slope and the CME may differ considerably, as Appendix A.1 and the examples in \citet{serranoserrat2026kernel} show.

The kernel estimator remains usable once its local regression is allowed to be nonlinear in $D$. The estimate at $X = x$ is then the average of each unit's fitted slope at its own treatment level, taken over units with that value of $X$. \citet{serranoserrat2026kernel} uses a similar averaging step after fitting a GAM. Appendix A.3 illustrates this modified estimator in one of that paper's examples.

With many covariates, we recommend a DML estimator with an orthogonal score for the average derivative. For a continuous treatment, the PO-Lasso and DML estimators in \citet{LLX} instead assume that the outcome is linear in $D$ with a slope common across $Z$. Otherwise they recover a variance-weighted linear projection. It averages the projection slopes of $Y$ on $D$ across values of $Z$, with more weight where $D$ varies more, and it need not equal the CME.

\paragraph*{Summary.} Contrary to the claim in \citet{simonsohn2024blog} that ``[A]ll three currently used estimators detect a spurious change in marginal effect,'' we demonstrate that the monotonically increasing marginal effect (or the CME) is not spurious, and the kernel estimator accurately recovers the true CME in this example, and the binning estimator remains a useful diagnostic tool. The confusion arises precisely because the causal estimand, CME, is not clearly defined in \citet{simonsohn2024blog} and is potentially conflated with either the CAPE or causal moderation. The claim that the kernel estimator fails to recover the correct answer is incorrect because he used the wrong benchmark. The kernel estimator was designed to identify the CME; when evaluated against the correct estimand, it produces the correct result in this example, whereas the GAM approach he proposed does not (see Figure~\ref{fig:sim}).

\bigskip

\section*{Advancements in Methods for Estimating the CME}

\citet{simonsohn2024blog} advocates for using GAM to estimate potentially nonlinear conditional relationships. While GAM is an important and easy-to-use exploratory tool, we do not consider it an appealing estimation strategy for the CME when there are many covariates in light of recent advancements in the causal inference literature. These advancements include (i) a shift in focus toward treatment assignment mechanisms (or ``designs'') and (ii) the development of robust estimation methods for targeted parameters with relaxed functional form assumptions, such as the introduction of double/debiased machine learning (DML) techniques  \citep[e.g.,][]{van2011targeted, chernozhukov2018double, athey2019generalized}. See also \citet{imbens2025comparing} for a recent survey of methods based on the unconfoundedness assumption.

We will explain the shortcomings of GAM in the next section. In this section, we briefly explain these advancements in the cross-sectional setting under the unconfoundedness and overlap assumptions.\footnote{More detailed discussions are in \citet{LLX}.} We also illustrate these estimators in a simulated example. First, we introduce two key identifying assumptions. 
\begin{assumption}\label{assm:unconf}
    (Unconfoundedness). 
\begin{center}
$Y_i(d) \perp\!\!\!\perp D_i \mid V_i = v, \text{ for all } d \in \mathcal{D} \text{ and } v \in \mathcal{V}$.
\end{center}
\end{assumption}

\begin{assumption} (Overlap)\label{assm:overlap}
$$f_{D \mid V}(d \mid v) > 0 \quad \text{for all } d \in \mathcal{D} \text{ and } v \in \mathcal{V}$$
where $\mathcal{D}$ is the support of the treatment $D$, $\mathcal{V}$ is the support of the covariates $V$, and $f_{D \mid V}(d \mid v)$ is the conditional probability density function of $D$ given $V$. When $D$ is binary, we require strict overlap: For some positive $\eta$, 
$$\eta \leq \mathbb{P}(D_i = 1 \mid V_i = v) \leq 1-\eta,\quad\text{with probability } 1.$$  
\end{assumption}
\noindent It means that, given the covariate values $V_i = v$, the probability of receiving the treatment is strictly bounded away from $0$ and $1$.

Under Assumptions~\ref{assm:sutva}-\ref{assm:overlap}, the CME is nonparametrically identified. However, when some of these variables are continuous or when $Z$ is high-dimensional, estimation requires additional dimension reduction tools. These tools may include specifying an outcome model or a model for the propensity score to facilitate estimation.

\begin{assumption} (Partially linear regression model, PLRM; \citealt{robinson1988root})\label{assm:plm}
\begin{align*}
& Y_i(d) = \theta(X_i)\, d + g(V_i) + \zeta_i, \quad \E[\zeta_i \mid D_i, V_i] = 0, \\
& D_i = m(V_i) + \varepsilon_i, \quad \mathbb{E}(\varepsilon_i \mid V_i) = 0.
\end{align*}
in which $g(\cdot)$ and $m(\cdot)$ are flexible functions. 
\end{assumption}
Note that when $D$ is binary, $m(V) = \pi(V)$ is the propensity score. The conditional mean restriction on $\zeta$ is part of the PLRM; unconfoundedness alone does not impose a slope common across $Z$. 

\paragraph*{DML and doubly robust estimators.} \citet{semenova2021debiased} show that under Assumptions~\ref{assm:sutva}-\ref{assm:overlap} and regularity conditions---for example, if the nuisance functions can be consistently estimated with a sufficiently fast convergence rate---the DML estimators, which satisfy the so-called Neyman orthogonality condition, can recover the CME and achieve consistency asymptotically. For binary $D$, AIPW-Lasso and AIPW-based DML use the propensity score $\pi(V)$ and the outcome regressions $\mu_d(V) = \E[Y \mid D = d, V]$; their score is doubly robust and Neyman orthogonal. For continuous $D$, the partially linear procedures in \citet{LLX} use $\ell(V) = \E[Y \mid V]$ and $m(V)$ and add the PLRM; their Robinson score is Neyman orthogonal for the variance-weighted linear projection, which equals the CME under the PLRM.

The primary strength of DML estimators lies in their robustness to model misspecification, as they rely on flexible machine learning algorithms to estimate nuisance functions, e.g., $\ell(V)$ and $m(V)$. This flexibility enables them to handle high-dimensional covariates and complex relationships between variables. Moreover, the use of Neyman orthogonality conditions ensures that small errors in estimating these nuisance functions do not propagate to the final estimate of the CME to first order, thereby enhancing robustness. 

The PLRM is general. It is straightforward to see that Model~(\ref{mod1}), the traditional parametric linear interaction model, and Model~(\ref{mod2}), the semiparametric kernel estimator proposed in HMX (2019), are both special cases of the PLRM. However, both models are limited by their restrictions on functional form. While the kernel estimator significantly relaxes the functional form assumptions of Model~(\ref{mod1}), it cannot accommodate complex, nonlinear relationships between $Z$ and $Y$. In other words, jointly modeling a complex response surface of $Y$ given $D$ and multiple $Z$ using the kernel estimator is either computationally infeasible or leads to significant regularization bias, affecting the estimation of target parameters in finite samples. The starting point of HMX (2019) is the assumption that the relationship between $Z$ and $Y$ is correctly specified, which allows us to focus on the nonlinearity of the CME and overlap issues. The PLRM relaxes this assumption.

\begin{figure}[!ht]
     \vspace{0.5cm}
     \caption{Estimating the CME with Various Estimators}
     \label{fig:sim.dml}
\centering
\begin{minipage}{1\linewidth}
\begin{center}
     \includegraphics[width=1\linewidth]{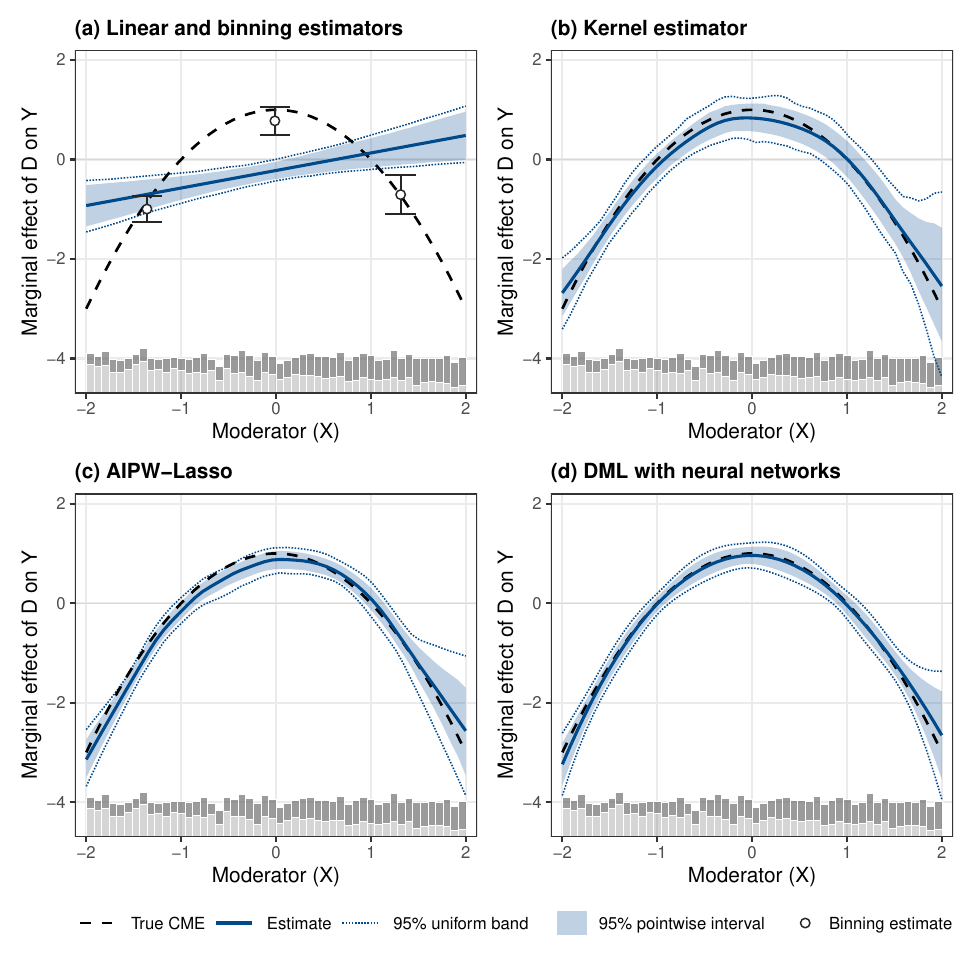}
\end{center}\vspace{-1em}
{\footnotesize \textbf{Note:} In each panel, the dashed line and the solid line represent the true and estimated CME, respectively. The shaded areas and the dotted lines depict the pointwise confidence intervals and uniform confidence bands, respectively. In panel (a), circles with bars are the binning estimates and their 95\% confidence intervals. At the bottom of each panel, a stacked histogram displays the distribution of treated (dark gray) and control (light gray) units across values of the moderator $X$. The sample size is 5,000. The moderator $X$ and additional covariates $Z_1$ and $Z_2$ enter the outcome model and $X$ and $Z_{1}$ enter the propensity score. The exact DGP can be found in Appendix Section A.2.} 
\end{minipage}
\end{figure}

The main limitations of DML estimators are twofold: (i) they are computationally intensive due to the need for cross-fitting and fine-tuning machine learning algorithms; and (ii) they require large sample sizes for reliable performance, as their effectiveness depends on accurately estimating nuisance functions. When data are scarce, DML estimators may not perform well. In such cases, doubly robust or orthogonal estimators based on parametric or semiparametric models, such as AIPW-Lasso \citep{robins1994estimation} and PO-Lasso \citep{belloni2014high}, offer reliable alternatives. \citet{blackwell2022reducing} provide an implementation of post-double-selection Lasso (PDS-Lasso) for estimating interaction effects, focusing on missing interactions between $X$ and $Z$ as well as $D$ and $Z$. The so-called \emph{fully moderated model} is also implemented in the \texttt{interflex} package. Building on Simonsohn's critique, we can extend the model by incorporating greater flexibility in modeling potential nonlinearity in $D$ and $V$ (including $X$). 

In Figure~\ref{fig:sim.dml}, we illustrate the performance of various estimators using a simulated example where the treatment is binary. The DGP is designed so that the moderator $X$ and two covariates, $Z_1$ and $Z_2$, enter the outcome equation in complex ways, and $X$ and $Z_1$ enter treatment assignment. The figure displays results for the linear and binning estimators (top-left), the kernel estimator (top-right), the AIPW-Lasso estimator (bottom-left), and the DML-NN estimator (bottom-right). Figure~\ref{fig:sim.dml} shows that the kernel, AIPW-Lasso, and DML estimators track the true CME across the moderator range in this draw, though less precisely near its edges, while the binning estimator remains a useful diagnostic tool. We do not include GAM because, with several covariates entering in complex ways, a GAM requires the researcher to choose which interactions to model; we detail the reasons in the next section.

\section*{\texorpdfstring{What GAM Estimates, and Its Limits}{What GAM Estimates, and Its Limits}} 

GAM has drawbacks compared to DML and other doubly robust or orthogonal estimators specifically designed for estimating the CME. GAM estimates the response surface, i.e., the conditional expectation of the \emph{outcome}, so the CME has to be computed from it in a second step. First, in practice, \citet{simonsohn2024blog} proposes the following procedure: (i) fit a GAM to model \( Y = f(D, X) + \epsilon \); (ii) estimate ``marginal effects'' at \( D = d_{0} \) using the difference in predicted values from the estimated model \( \hat{f} \):
\[
\widehat{\Delta}_{h}(d_{0}, x) = \frac{\hat{f}(d_{0} + h, x) - \hat{f}(d_{0}, x)}{h}, \qquad h = 0.1.
\]
For differentiable $f$, this estimates the average slope over $[d_0, d_0 + h]$, $\Delta_h(d_0, x) = \{f(d_0 + h, x) - f(d_0, x)\}/h$, which approaches $\partial f(d, x)/\partial d$ at $d = d_0$ as $h \to 0$; with no additional covariates and under unconfoundedness, this is the CAPE $\rho(d_0, x) = \E[\partial Y_i(d)/\partial d|_{d = d_0} \mid D_i = d_0, X_i = x]$. This approach approximates the CAPE instead of the CME, because it evaluates the slope at a single treatment level $d_0$ rather than averaging each unit's slope at its own treatment level over units with $X = x$. Averaging those unit-level slopes would recover the CME; with no covariates and a continuous treatment that is not normal given $X$, such a GAM can outperform the kernel estimator \citep{serranoserrat2026kernel}. Moreover, the CAPE estimate depends on the choice of $d_0$, which introduces researcher discretion.

Second, as a spline-based semiparametric estimator, GAM requires users to specify particular nonlinear relationships to include in the model, making it prone to model misspecification. When additional covariates are present, manually selecting which nonlinear relationships to focus on becomes impractical. As demonstrated by \citet{hainmueller2014kernel}, GAM is not a reliable conditioning strategy when multiple covariates need to be modeled, such as in the case of $g(V)$. In other words, jointly modeling a complex response surface of $Y$ given $D$ and multiple $V$ quickly becomes impractical (a tensor-product smooth in more than a few variables needs many coefficients, so in practice only a few interactions can be modeled).

Third, unlike DML estimators, GAM does not satisfy the Neyman orthogonality condition, which prevents small errors in estimating nuisance functions from propagating to the final estimates to first order. As a result, it suffers from regularization bias due to its reliance on penalized splines, particularly in high-dimensional, data-rich settings. Similarly, because GAM does not model the treatment assignment mechanism, it lacks the double robustness of AIPW-Lasso and the orthogonality of PO-Lasso, making it a less appealing choice in small-to-medium sample size scenarios.   
 
\begin{figure}[!h]
     \caption{Kernel Estimates of the CME and GAM Estimates of Fixed-Dose Contrasts}\label{fig:sim.gam}
     \begin{minipage}{\linewidth}
     \begin{center}
     \includegraphics[width=0.8\linewidth]{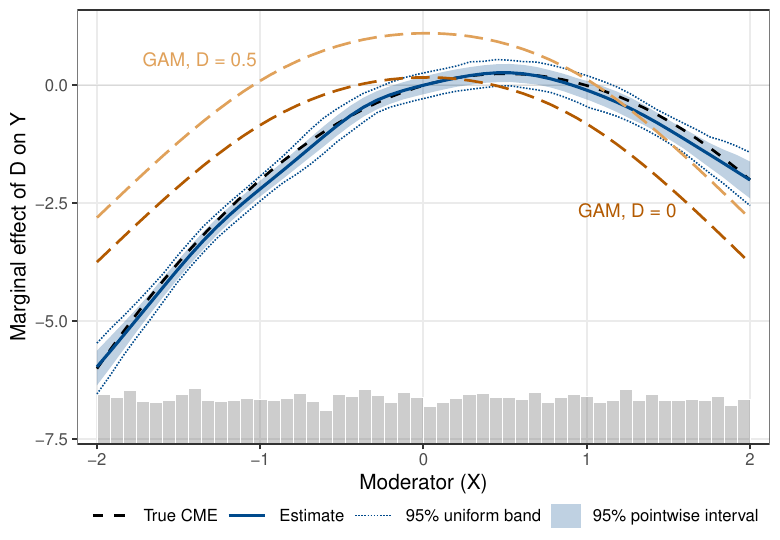}
     \end{center}\vspace{-1em}
     \label{fig:continuous_gam}
     {\footnotesize \textbf{Note:} The dashed line and the solid line represent the true CME, $\theta(x) = x - x^2$, and estimated CME using the kernel method, respectively. The shaded area and the dotted lines depict the pointwise confidence intervals and uniform confidence bands, respectively. The histogram at the bottom displays the distribution of units across different values of $X$. In addition, the dark and light orange long-dashed lines depict GAM estimates of the fixed-dose contrasts $\Delta_{0.1}(0, x) = 0.1 - x^2$ and $\Delta_{0.1}(0.5, x) = 1.1 - x^2$, which approximate the CAPEs $\rho(0, x) = -x^2$ and $\rho(0.5, x) = 1 - x^2$. The kernel estimator recovers the CME here because $D$ is normal given $X$.}     
    \end{minipage}
\end{figure}

Below, we demonstrate the first problem: the GAM output at a fixed treatment level approximates the CAPE, not the CME. We simulate a dataset with a continuous treatment $D$, a continuous moderator $X$, and no additional covariates. The exact DGP is provided in Appendix Section A.2. The true CME is $\theta(x) = x - x^2$. However, the fixed-dose contrast computed from a GAM fit with the \texttt{mgcv} package in \texttt{R}---as in the blog's figure \citep{simonsohn2024blog}---does not estimate the CME but instead approximates the conditional average partial effect (CAPE), $\rho(d, x)$.  As shown in Figure~\ref{fig:sim.gam}, $\rho(d, x)$ varies drastically depending on $d$, the level of $D$ being conditioned on, which can lead to significant confusion in applied research. This issue highlights the importance of clearly defining the estimand from the outset to avoid misinterpretation.

\section*{Conclusion and Recommendations}

Recent critiques of HMX (2019) reignite important discussions about robustly estimating conditional causal relationships. However, these critiques have critical flaws. First, the lack of a well-defined estimand creates confusion about the theoretical target, leading to erroneous conclusions (such as the claim that the kernel estimator cannot recover the quantity of interest in the key simulated example). Second, the main recommendation---using GAM---as implemented approximates the CAPE at a chosen treatment level rather than the CME, and falls short in addressing key empirical challenges, particularly when additional covariates are present. Building on HMX (2019) and subsequent works, we propose the following recommendations:
\begin{itemize}
    \item Clearly state the quantity of interest and the key identifying and modeling assumptions before proceeding with the analysis.
    \item Assess data quality by checking for missingness and outliers, and evaluate whether the overlap assumption is satisfied; improve overlap by trimming the data if necessary. 
    \item Use GAM, the binning estimator, and the kernel estimator as exploratory and diagnostic tools to understand potential nonlinear relationships.
    \item Adopt flexible modeling strategies. The kernel estimator lets the effect vary nonlinearly with $X$; DML and other doubly robust or orthogonal estimators also handle many covariates.\\ \\
    {\small\onehalfspacing\setlength{\extrarowheight}{8pt}
    \begin{tabular}{>{\raggedright\arraybackslash}p{0.45\linewidth}|>{\raggedright\arraybackslash}p{0.45\linewidth}}\hline\hline
     Setting    &  Suitable Estimators\\ \hline
    1) Experimental, binary $D$  & Kernel estimator \\
    2) Observational w/ a small-to-medium $n$ or smooth nuisance functions & AIPW-Lasso or PO-Lasso \\
    3) Observational w/ a large $n$ and complex nuisance functions & DML with flexible learners (NN, RF, HGB) \\
    4) Continuous $D$, outcome possibly nonlinear in $D$ & With few covariates, average each unit's fitted slope over units at each $X$, after a GAM fit or the kernel fit (Appendix A.3); otherwise a DML estimator built on an average-derivative score \\ \hline 
    \end{tabular}}\\ \\
    {\onehalfspacing\textbf{Note:} The doubly robust and DML estimators target the same quantity as the kernel estimator when no additional covariates exist or confound the relationship between $D$ and $Y$. In addition, the kernel estimator remains a valid tool when the covariates enter the outcome model linearly and, for a continuous treatment, the outcome is linear in $D$. With no additional covariates, the projection slope that the kernel estimator converges to equals the CME for every sufficiently regular dose-response function if and only if $D$ is normal given $X$; random assignment alone does not guarantee this equality. For a continuous treatment, the PO-Lasso and DML estimators in the second and third rows assume an outcome that is linear in $D$; the last row covers outcomes that may be nonlinear in $D$. The DML, AIPW-Lasso and PO-Lasso estimators are implemented in the \texttt{interflex} package \citep{LLX}.}
    \item Interpret results carefully. Use causal language with caution. For example, CME describes the causal effect of $D$ along the moderator $X$, but it does not represent the causal effect of $X$ itself.
\end{itemize}
Finally, we thank Professor Simonsohn again for the constructive criticism that inspired this response. 

\vspace{3em}
\FloatBarrier
\singlespacing
\bibliographystyle{apsr}
\bibliography{ref.bib}
\clearpage

\appendix
\onehalfspacing
\setcounter{page}{1}
\setcounter{table}{0}
\setcounter{figure}{0}
\setcounter{equation}{0}
\setcounter{footnote}{0}
\renewcommand{\theassumption}{A\arabic{assumption}}
\renewcommand\thetable{A\arabic{table}}
\renewcommand\thefigure{A\arabic{figure}}
\renewcommand{\thepage}{A-\arabic{page}}
\renewcommand{\theequation}{A\arabic{equation}}
\renewcommand{\thefootnote}{A\arabic{footnote}}

\vspace{0em}
\section{Appendix}
\bigskip

\subsection{Dissecting the Key Simulated Example}

We begin by deriving the true CME. We then show that the linear interaction model is biased, while the kernel estimator proposed in HMX (2019) can recover the CME when data are abundant, because $D$ is normal given $X$. 

\paragraph*{The true CME.} The partial derivative of $Y$ with respect to $D$ is
\[
\frac{\partial Y}{\partial D} \;=\; 2D \;-\; 0.5.
\]
To obtain the conditional marginal effect at $X=x$, we compute
\[
\E\!\Bigl[\frac{\partial Y}{\partial D} \Bigm| X=x\Bigr]
\;=\;
\E[\,2D - 0.5 \,\mid X=x].
\]
Under the bivariate normal assumption $\operatorname{Corr}(D,X) = 0.5$, it follows that
\[
\E[D \mid X=x] = \E[D] + \frac{\mathrm{Cov}(D,X)}{\mathrm{Var}(X)}(x - \E[X]) =  0.5 x.
\]
Therefore, the true CME is
\[
\theta(x) 
\;=\;
\E[\,2D - 0.5 \mid X=x]
\;=\;
2\bigl(0.5\,x\bigr) - 0.5
\;=\;
x - 0.5.
\]

\paragraph*{The linear interaction model is biased for CME.} We now consider the misspecified linear--interaction model:
\[
Y 
\;=\;
\beta_0 \;+\; \beta_d\,D \;+\; \beta_x\,X 
\;+\; \beta_{dx}\,(D \times X) 
\;+\; \text{error}.
\]
Under our normality assumptions, the $4 \times 4$ matrix 
$\displaystyle
M = \E\bigl[\mathbf{Z}\,\mathbf{Z}^\top\bigr]$,
where $\mathbf{Z} =(1\ D\ X\ D\,X)^T$,
together with 
$\displaystyle
\mathbf{b} = \E[\mathbf{Z}\,Y],
$
fully determines the population OLS solution:
\[
\beta 
\;=\; 
\bigl(\beta_0,\;\beta_d,\;\beta_x,\;\beta_{dx}\bigr)^\top 
\;=\;
M^{-1}\,\mathbf{b}.
\]
Carrying out the matrix calculations, we find
\[
\beta_0 = 0.6,\quad \beta_d = -0.5,\quad \beta_x = 0,\quad \beta_{dx} = 0.8.
\]
Thus, the fitted regression implies the conditional marginal effect
\[
\widehat{\text{CME}}(X) 
\;=\;
\frac{\partial \hat{Y}}{\partial D}
\;=\;
\beta_d \;+\; \beta_{dx}\,X
\;=\;
-0.5 \;+\; 0.8\,X.
\]
Note that this differs from the true CME, $X - 0.5$. 

\paragraph*{The kernel estimator can recover the CME.}  Although the true DGP features a $D^2$ term, the kernel estimator can recover the true CME, that is, $\theta(x) = x - 0.5$  (in regions where data are abundant).  To understand why, consider the following local‐linear setup. Define $f(x)$ and $g(x)$ as the intercept and slope of the best linear projection of $Y$ on $D$ among units with $X = x$:
\[
  Y \;=\; f(X) \;+\; g(X)\,D \;+\; \epsilon,
  \quad
  \E[\epsilon \mid X] = \E[D\epsilon \mid X] = 0,
\]
where $f$ and $g$ are unknown functions of $X$ that we approximate \emph{locally} around $X = x_0$ using linear functions.  
Specifically, for $\lvert X - x_0\rvert < r$ (a small neighborhood), write
\[
  f(X)_{\lvert X - x_0\rvert < r} 
  \;\approx\; 
  \mu(x_0) + \eta(x_0)\,\bigl(X - x_0\bigr),
  \quad
  g(X)_{\lvert X - x_0\rvert < r}
  \;\approx\;
  \alpha(x_0) + \beta(x_0)\,\bigl(X - x_0\bigr).
\]
We then estimate the local coefficients $\bigl\{\mu(x_0),\,\eta(x_0),\,\alpha(x_0),\,\beta(x_0)\bigr\}$ by solving the following weighted least‐squares problem:
\[
\begin{aligned}
&\bigl(\hat{\mu}(x_0),\,\hat{\alpha}(x_0),\,\hat{\eta}(x_0),\,\hat{\beta}(x_0)\bigr)
\;=\;
\arg\min_{\tilde{\mu},\,\tilde{\alpha},\,\tilde{\eta},\,\tilde{\beta}}
\;
L\bigl(\tilde{\mu},\,\tilde{\alpha},\,\tilde{\eta},\,\tilde{\beta}\bigr),\\[6pt]
&L\;=\;
\sum_{i=1}^{N} 
\Bigl[
  Y_i 
  \;-\;
  \tilde{\mu}
  \;-\;
  \tilde{\alpha}\,D_i
  \;-\;
  \tilde{\eta}\,\bigl(X_i - x_0\bigr)
  \;-\;
  \tilde{\beta}\,\bigl[D_i\,(X_i - x_0)\bigr]
\Bigr]^{2}
\;
K\!\Bigl(\,\frac{X_i - x_0}{h(x_0)}\Bigr),
\end{aligned}
\]
where $K\!\bigl(\cdot\bigr)$ is a \emph{kernel weight} function, and $h(x_0)$ is the \emph{bandwidth} controlling how quickly weights decay as $X_i$ moves away from $x_0$.  

For simplicity, consider the uniform kernel $K(z) = 1$, if $|z| < 1$ and $0$ otherwise. As $h(x_0)\to 0$, this kernel enforces $\lvert X_i - x_0\rvert < h(x_0)$ to be very small, so effectively only points with $X_i \approx x_0$ remain in the sum.  With $nh(x_0) \to \infty$, a positive density of $X$ and $\mathrm{Var}(D \mid X) > 0$ at $x_0$, and conditional moments continuous in $x$, the coefficient on $D$ converges to the slope of the best linear projection of $Y$ on $D$ among units with $X = x_0$. In the neighborhood of $x_0$, this slope is $\widehat{\mathrm{Cov}}(D,Y)$ divided by $\widehat{\mathrm{Var}}(D)$ (conditional on $X\approx x_0$), and bivariate normality with $\mathrm{Cov}(D,X)=0.5$ gives its value:
\[
  \hat{\alpha}(x_0)
  \;\approx\;
  \frac{\mathrm{Cov}(D,Y \,\vert\,X\approx x_0)}
       {\mathrm{Var}(D \,\vert\,X\approx x_0)}
  \;\approx\;
  x_0 \;-\; 0.5.
\]
Thus, $\hat{\alpha}(x_0)$ converges to the \emph{true} conditional marginal effect $\theta(x_0)= x_0 - 0.5$. The last step uses normality. If $D$ is normal given $X$, then $\mathrm{Cov}(D, D^2 \mid X) = 2\,\E[D \mid X]\,\mathrm{Var}(D \mid X)$ (Stein's lemma), so the projection slope equals $\E[2D - 0.5 \mid X = x_0] = \theta(x_0)$. For this quadratic outcome, the projection slope minus the CME equals $\E[(D - \mu_{x_0})^3 \mid X = x_0]/\mathrm{Var}(D \mid X = x_0)$, where $\mu_{x_0} = \E[D \mid X = x_0]$, so equality holds whenever the conditional third central moment is zero; for every sufficiently regular outcome function, it requires normality (see below).

\paragraph*{What the kernel estimator recovers in general.} Consider $Y = s(D, X) + \epsilon$ with $s$ smooth, $\E[\epsilon \mid D, X] = 0$, and $D$ continuous. As $h \to 0$ and $nh \to \infty$, the slope of the local regression at $x_0$ converges to the slope of the best linear projection of $Y$ on $D$ among units with $X = x_0$, so
\[
\hat{\alpha}(x_0) \xrightarrow{p} \frac{\mathrm{Cov}(D, Y \mid X = x_0)}{\mathrm{Var}(D \mid X = x_0)} = \int \frac{\partial s(d, x_0)}{\partial d}\, w(d \mid x_0)\, \mathrm{d}d,
\]
where $w(d \mid x_0) = \bigl\{\E[D \mid D > d, X = x_0] - \E[D \mid X = x_0]\bigr\}\bigl\{1 - F(d \mid x_0)\bigr\} / \mathrm{Var}(D \mid X = x_0)$ are non-negative weights that integrate to one \citep{yitzhaki1996using}. The CME instead weights $\partial s(d, x_0)/\partial d$ by the density $f_{D \mid X}(d \mid x_0)$. The two coincide when $s$ is linear in $D$, and they coincide for every $s$ only when $w = f_{D \mid X}$, that is, when $D$ is normal given $X = x_0$ \citep{serranoserrat2026kernel}. The February 2025 version of this response claimed that $\hat{\alpha}(x_0)$ converges to the CME for any smooth $s$. That claim was incorrect: a regression that is local in $X$ is still a single linear regression in $D$, so its slope averages $\partial s/\partial d$ with the weights $w$ rather than with the density of $D$.

\clearpage

\subsection{DGP for the Simulated Examples}

\paragraph*{DGP for the simulated example in Figure~\ref{fig:sim.dml}.}

The true DGP is defined as follows: the moderator \( X \) is uniformly distributed on \([-2, 2]\), the covariates \( Z_1 \) and \( Z_2 \) are independent standard normal variables. The treatment assignment is given by:
\[
\P(D = 1 \mid V) = \frac{e^{\left(0.5 X+0.5 Z_1\right)}}{1+e^{\left(0.5 X+0.5 Z_1\right)}}.
\] 
The outcome \( Y \) is generated according to
\[
Y = 1 + X^2 + D - X^2 D + e^{Z_1 + 0.5 X} Z_2 + Z_1^2 - 2 \cdot \mathbb{I}(Z_2 > 0) Z_2 + 3 \sin(Z_1 + Z_2) + \epsilon,
\]
where \( \epsilon \sim \mathcal{N}(0,1) \). The true CME is $\theta(x) = 1 - x^2$. $X$, $Z_1$, $Z_2$ and $\epsilon$ are drawn independently, and $\epsilon$ is independent of treatment assignment. This design satisfies pointwise positivity, $0 < \P(D = 1 \mid V) < 1$, but because $Z_1$ is unbounded it does not satisfy the uniform bound in Assumption~\ref{assm:overlap}; Figure~\ref{fig:sim.dml} is an illustration under pointwise positivity.

\paragraph*{DGP for the simulated example in Figure~\ref{fig:sim.gam}.}

The true DGP is defined as follows: the moderator \( X \) is uniformly distributed on \([-2, 2]\). The treatment variable \( D \) is generated by
\[ D = 0.5 X + \epsilon_D, \]
where \( \epsilon_D \sim \mathcal{N}(0,1) \), drawn independently of $X$. The outcome \( Y \) is generated according to
\[
Y = 1 + 1.5 X + D^2 - D X^2 + \epsilon_Y,
\] 
where \( \epsilon_Y \sim \mathcal{N}(0,1) \), drawn independently of $X$ and $\epsilon_D$. The true CME is 
\[ \E\left(\frac{\partial Y}{\partial D} \Big| X  = x\right) = x - x^2. \] 
\clearpage

\subsection{A Kernel Estimator Flexible in the Treatment: An Illustration}

The discrepancy described at the end of Appendix A.1 arises not from kernel smoothing in $X$ but from fitting a straight line in $D$ at each value of $X$. A simple fix keeps the kernel weighting in $X$ but lets the outcome be nonlinear in $D$. At each evaluation point $x_0$, we regress $Y$ on a cubic B-spline basis in $D$ (five degrees of freedom), each term interacted with $(1, X - x_0)$, using the standard estimator's Gaussian kernel weights, with the single bandwidth $h$ that \texttt{interflex} selects by cross-validation for that estimator. Each unit's slope is the derivative of the curve fitted at $x_0$, evaluated at the unit's own treatment level, and $\hat\theta(x_0)$ is the intercept of a kernel-weighted local linear regression of these slopes on $X - x_0$. This is a plug-in estimator: it first estimates the outcome surface, then averages its derivative. \citet{serranoserrat2026kernel} uses a similar averaging step after fitting a GAM; here we apply it to the kernel estimator after letting its local regression be flexible in $D$.

\begin{figure}[H]
    \caption{Kernel Estimates in the Gaussian-Copula Design}\label{fig:flexD}
    \centering
    \begin{minipage}{0.97\linewidth}
    \begin{center}
    \includegraphics[width=1\linewidth]{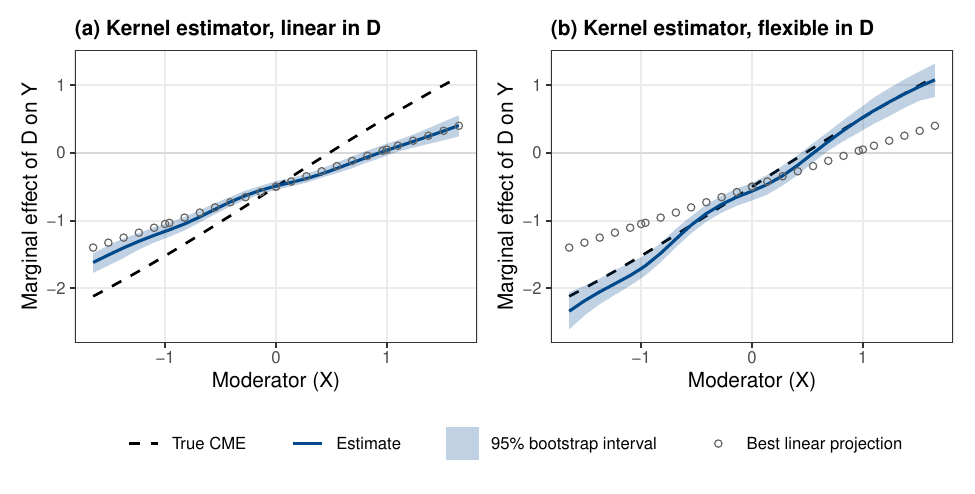}
    \end{center}\vspace{-1em}
    {\footnotesize \textbf{Note:} One draw with $N = 5{,}000$ from the Gaussian-copula design of \citet{serranoserrat2026kernel}. Solid lines are estimates and shaded areas are 95\% pointwise percentile bootstrap intervals (200 resamples, bandwidth held fixed). The dashed line is the true CME; circles mark the slope of the best linear projection of $Y$ on $D$ at each value of $X$. Panel (a): the kernel estimator in \texttt{interflex} (point estimates). Panel (b): the same kernel weights and bandwidth, with a cubic B-spline in $D$ and unit-level slopes averaged by a local linear fit in $X$.}
    \end{minipage}
\end{figure}

Figure~\ref{fig:flexD} applies both estimators to one draw ($N = 5{,}000$) of the Gaussian-copula design in \citet{serranoserrat2026kernel}: $Y = D^2 - 0.5D + \varepsilon$, with a treatment that is uniform overall but not given $X$. The standard kernel estimator (panel a) follows the projection slope over most of the range rather than the CME. At $X = 1$ it gives 0.04, with a 95\% interval of $[-0.04, 0.13]$, while the CME is 0.52. Once the local regression is flexible in $D$ (panel b), the estimate moves from the projection slope to the CME: 0.52 at $X = 1$, with an interval of $[0.34, 0.69]$. In a single draw the fit is not exact everywhere: at $X = -1$ the estimate is $-1.71$, with an interval of $[-1.85, -1.54]$, against a CME of $-1.52$; the standard estimator gives $-1.16$ there. Kernel smoothing in $X$ is not the problem: once the local regression accounts for nonlinearity in $D$, the estimator targets the CME.

This fix is an illustration, not a general solution. It works here, with no covariates and a smooth outcome, but errors in the fitted curve pass into the estimate to first order, and a flexible surface in $D$ and many covariates is impractical to fit. With many covariates, the principled solution is a DML estimator built on a Neyman-orthogonal score for the average derivative, such as the density-based score of \citet{semenova2021debiased} or the automatic debiasing of \citet{chernozhukov2022automatic}, with the debiased unit-level slopes then projected onto a basis in $X$. The partially linear DML and PO-Lasso estimators currently in \texttt{interflex} instead target the variance-weighted linear projection described in the main text.

\end{document}

%% file: preamble.tex
\usepackage[margin=1in]{geometry}

\usepackage{fullpage}
\usepackage[utf8]{inputenc}
\usepackage[english]{babel}
\usepackage[shortlabels]{enumitem}
\usepackage{mathtools,
  booktabs,
  braket,
  mathrsfs,
  latexsym,
  epsfig,
  xcolor,
  url,
  setspace,
  graphics,
  lipsum,
  float,
  lineno,
  placeins
}
\usepackage{subfigure}
\usepackage{float}
\usepackage{pdflscape}
\newcommand{\bland}{\begin{landscape}}
\newcommand{\eland}{\end{landscape}}

\usepackage[flushmargin]{footmisc}
\definecolor{CardinalRed}{cmyk}{0,1,0.65,0.34} %
\usepackage[colorlinks,linkcolor=CardinalRed,citecolor=CardinalRed,urlcolor = CardinalRed]{hyperref}
\usepackage{hyperref}

\makeatletter
\def\maxwidth{\ifdim\Gin@nat@width>\linewidth\linewidth\else\Gin@nat@width\fi}
\def\maxheight{\ifdim\Gin@nat@height>\textheight\textheight\else\Gin@nat@height\fi}
\makeatother

\setkeys{Gin}{width=\maxwidth,height=\maxheight,keepaspectratio}

\newcommand{\burl}[1]{\textcolor{blue}{\url{#1}}}

\usepackage[round]{natbib}

\usepackage[OT1]{fontenc}

\usepackage{todonotes}

\makeatletter
\providecommand\@dotsep{5}
\def\listtodoname{List of Todos}
\def\listoftodos{\@starttoc{tdo}\listtodoname}
\makeatother

\graphicspath{{./graphs/}}%

\usepackage[labelsep=period, font=sc, skip=0pt,justification=centering]{caption}
\usepackage[figuresright]{rotating}
\usepackage{titlesec}
\titlelabel{\thetitle.\ }

\usepackage{titling}
\usepackage{titlesec}
\titleformat{\section}
{\normalfont\fontsize{15}{15}\bfseries}{\thesection.}{0.5em}{}

\newenvironment{itemize*}%
  {\begin{itemize}%
    \setlength{\itemsep}{0pt}%
    \setlength{\parskip}{0pt}}%
  {\end{itemize}}
\newenvironment{enumerate*}%
  {\begin{enumerate}%
    \setlength{\itemsep}{0pt}%
    \setlength{\parskip}{0pt}}%
  {\end{enumerate}}

\usepackage{
  amscd,
  amsfonts,
  amsmath,
  amssymb,
  amsbsy,
  amsthm,
  bm, %
  dsfont,
  cancel, %
  latexsym,
  mathtools,
  graphicx,
  xcolor,
  xargs
}

\usepackage{longtable}

\newcommand{\beq}{\begin{equation}}
\newcommand{\eeq}{\end{equation}}

\newcommand*\Bigpar[1]{\left( #1 \right )}

\newcommand{\ba}{\begin{array}}
\newcommand{\ea}{\end{array}}
\newcommand{\be}{\begin{enumerate}}
\newcommand{\ee}{\end{enumerate}}
\newcommand{\bi}{\begin{itemize}}
\newcommand{\ei}{\end{itemize}}
\newcommand{\bs}{\begin{align}\begin{split}\nonumber}
\newcommand{\bsnumber}{\begin{align}\begin{split}}
\newcommand{\es}{\end{split}\end{align}}

\newcommandx{\deriv}[2][1=x,2=f]{\nabla \, #2 \Bigpar{ #1 } }

\renewcommand{\to}{{\rightarrow}}

\newcommand\frakfamily{\usefont{U}{yfrak}{m}{n}}
\DeclareTextFontCommand{\textfrak}{\frakfamily}

\renewcommand{\to}{{\rightarrow}}

\newcommand{\E}{\mathbb{E}} %
\renewcommand{\P}{\mathbb{P}} %

\newcommand{\hyp}[2]{
\ensuremath{H_0:#1 \ifhmode\quad\text{versus}\quad\fi\text{ vs. } H_1:#2}}

\newcommandx{\uniff}[1][1={a,b}]{\textrm{Unif}\left({#1}\right)}
\newcommandx{\unifd}[1][1={a,\ldots,b}]{\textrm{Unif}\left\{{#1}\right\}}
\newcommandx{\dunif}[3][1=x,2=a,3=b]{\frac{I(#2<#1<#3)}{#3-#2}}
\newcommandx{\dunifd}[3][1=x,2=a,3=b]{\frac{I(#2\le#1\le#3)}{#3-#2+1}}
\newcommandx{\punif}[3][1=x,2=a,3=b]{
\begin{cases} 0 & #1 < #2 \\ \frac{#1-#2}{#3-#2} & #2 < #1 < #3 \\ 1 & #1 > #3\\\end{cases}}
\newcommandx{\punifd}[3][1=x,2=a,3=b]{
\begin{cases} 0 & #1 < #2\\ \frac{\lfloor#1\rfloor-#2+1}{#3-#2} & #2 \le #1 \le #3 \\ 1 & #1 > #3\\ \end{cases}}

\newcommandx\bern[1][1=p]{\textrm{Bern}\left({#1}\right)}
\newcommandx\dbern[2][1=x,2=p]{#2^{#1} \left(1-#2\right)^{1-#1}}
\newcommandx\pbern[2][1=x,2=p]{\left(1-#2\right)^{1-#1}}

\newcommandx\bin[1][1={n,p}]{\textrm{Bin}\left(#1\right)}
\newcommandx\dbin[3][1=x,2=n,3=p]{\binom{#2}{#1}#3^#1\left(1-#3\right)^{#2-#1}}

\newcommandx\mult[1][1={n,p}]{\textrm{Mult}\left(#1\right)}
\newcommandx\dmult[3][1=x,2=n,3=p]{\frac{#2!}{#1_1!\ldots#1_k!}#3_1^{#1_1}\cdots#3_k^{#1_k}}

\newcommandx\hyper[1][1={N,m,n}]{\textrm{Hyp}\left({#1}\right)}
\newcommandx\dhyper[4][1=x,2=N,3=m,4=n]{\frac{\binom{#3}{#1}\binom{#2-#3}{#4-#1}}{\binom{#2}{#4}}}

\newcommandx\nbin[1][1={r,p}]{\textrm{NBin}\left({#1}\right)}
\newcommandx\dnbin[3][1=x,2=r,3=p]{\binom{#1+#2-1}{#2-1}#3^#2(1-#3)^#1}
\newcommandx\pnbin[3][1=x,2=r,3=p]{I_#3(#2,#1+1)}

\newcommandx\geo[1][1=p]{\textrm{Geo}\left(#1\right)}
\newcommandx\dgeo[2][1=x,2=p]{#2(1-#2)^{#1-1}}
\newcommandx\pgeo[2][1=x,2=p]{1-(1-#2)^#1}

\newcommandx\pois[1][1=\lambda]{\textrm{Po}\left({#1}\right)}
\newcommandx\dpois[2][1=x,2=\lambda]{\frac{#2^#1 e^{-#2}}{#1!}}
\newcommandx\ppois[2][1=x,2=\lambda]{e^{-#2}\sum_{i=0}^#1\frac{#2^i}{i!}}

\newcommandx\normall[1][1={\mu,\sigma^2}]{\mathcal{N}\left({#1}\right)}
\newcommandx\dnormall[3][1=x,2=\mu,3=\sigma]%
  {\frac{1}{#3\sqrt{2\pi}}\exp \Bigpar{-\frac{\left(#1-#2\right)^2}{2 #3^2}}}
\newcommandx\pnormall[1][1=x]{\Phi\left({#1}\right)}
\newcommandx\qnormall[1]{\Phi^{-1}\left({#1}\right)}

\newcommandx\mvn[1][1={\mu,\Sigma}]{\mathrm{MVN}\left({#1}\right)}

\newcommandx\ex[1][1=\lambda]{\textrm{Exp}\left(#1\right)}
\newcommandx\dex[2][1=x,2=\lambda]{#2e^{-#1 #2}}
\newcommandx\pex[2][1=x,2=\lambda]{1-e^{-#1 #2}}

\newcommandx\gam[1][1={\alpha,\lambda}]{\textrm{Gamma}\left({#1}\right)}
\newcommandx\dgamma[3][1=x,2=\alpha,3=\lambda]%
  {\frac{#3^{#2}}{\Gamma\left( #2 \right)} #1^{#2-1}e^{-#3#1}}

\newcommandx\invgamma[1][1={\alpha,\beta}]{\textrm{InvGamma}\left({#1}\right)}
\newcommandx\dinvgamma[3][1=x,2=\alpha,3=\beta]%
{\frac{#3^{#2}}{\Gamma\left(#2\right)}#1^{-#2-1}e^{-#3/#1}}
\newcommandx\pinvgamma[3][1=x,2=\alpha,3=\beta]%
{\frac{\Gamma\left(#2,\frac{#3}{#1}\right)}{\Gamma\left(#2\right)}}

\newcommandx\bet[1][1={\alpha,\beta}]{\textrm{Beta}\left(#1\right)}
\newcommandx\dbeta[3][1=x,2=\alpha,3=\beta]
{\frac{\Gamma\left(#2+#3\right)}{\Gamma\left(#2\right)\Gamma\left(#3\right)}#1^{#2-1}\left(1-#1\right)^{#3-1}}

\newcommandx\dir[1][1={\alpha}]{\textrm{Dir}\left(#1\right)}
\newcommandx\ddir[3][1=x,2=\alpha]{\frac{\Gamma\left(\sum_{i=1}^k #2_i\right)}{\prod_{i=1}^k\Gamma\left(#2_i\right)}\prod_{i=1}^k #1_i^{#2_i-1}}

\newcommandx\weibull[1][1={\alpha}]{\textrm{Dir}\left(#1\right)}
\newcommandx\dweibull[3][1=x,2=\lambda,3=k]{\frac{#3}{#2}
\left(\frac{#1}{#2}\right)^{#3-1} e^{-(#1/#2)^k}}

\newcommandx\chisq[1][1=k]{\chi_{#1}^2}

\newcommandx\zet[1][1=s]{\textrm{Zeta}\left(#1\right)}
\newcommandx\dzeta[2][1=x,2=s]{\frac{#1^{-#2}}{\zeta\left(#2\right)}}

\newtheoremstyle{mystyle}%
  {12pt}%
  {12pt}%
  {}%
  {}%
  {\sffamily \bfseries }%
  {.}%
  {0.5em}%
  {\thmname{#1}\thmnumber{ #2}\thmnote{ (#3)}}%

\theoremstyle{mystyle}

\newtheorem{assumption}{Assumption}

\newenvironment{proof-sketch}{\noindent{\bf Sketch of Proof}
  \hspace*{1em}}{\qed\bigskip\\}
\newenvironment{proof-idea}{\noindent{\bf Proof Idea}
  \hspace*{1em}}{\qed\bigskip\\}
\newenvironment{proof-of-lemma}[1][{}]{\noindent{\bf Proof of Lemma {#1}}
  \hspace*{1em}}{\qed\bigskip\\}
\newenvironment{proof-of-proposition}[1][{}]{\noindent{\bf
    Proof of Proposition {#1}}
  \hspace*{1em}}{\qed\bigskip\\}
\newenvironment{proof-of-theorem}[1][{}]{\noindent{\bf Proof of Theorem {#1}}
  \hspace*{1em}}{\qed\bigskip\\}
\newenvironment{inner-proof}{\noindent{\bf Proof}\hspace{1em}}{
  $\bigtriangledown$\medskip\\}
\newenvironment{proof-attempt}{\noindent{\bf Proof Attempt}
  \hspace*{1em}}{\qed\bigskip\\}

\usepackage{array}
\newcolumntype{L}[1]{>{\raggedright\let\newline\\\arraybackslash\hspace{0pt}}m{#1}}
\newcolumntype{C}[1]{>{\centering\let\newline\\\arraybackslash\hspace{0pt}}m{#1}}
\newcolumntype{R}[1]{>{\raggedleft\let\newline\\\arraybackslash\hspace{0pt}}m{#1}}

\usepackage{multirow}

\def\E{\mathbb{E}} %
\def\P{\mathbb{P}} %